\documentclass[prl,amsmath,amssymb,superscriptaddress,preprint]{revtex4}
\usepackage{graphicx}
\usepackage{dcolumn}
\usepackage{bm}
\usepackage{here}

\begin{document}

\title{Entanglement generation and multiparticle interferometry with neutral atoms.}

\author{Artem M. Dudarev}
\affiliation{Department of Physics, The University of Texas,
Austin, Texas 78712-1081} \affiliation{Center for Nonlinear
Dynamics, The University of Texas, Austin, Texas 78712-1081}
\author{Roberto B. Diener}
\affiliation{Department of Physics, The University of Texas,
Austin, Texas 78712-1081}
\author{Biao Wu}
\affiliation{Department of Physics, The University of Texas,
Austin, Texas 78712-1081} \affiliation{Solid State Division, Oak
Ridge National Laboratory, Oak Ridge, Tennessee 37831-6032}
\author{Mark G. Raizen}
\affiliation{Department of Physics, The University of Texas,
Austin, Texas 78712-1081} \affiliation{Center for Nonlinear
Dynamics, The University of Texas, Austin, Texas 78712-1081}
\author{Qian Niu}
\affiliation{Department of Physics, The University of Texas,
Austin, Texas 78712-1081}

\date{\today}

\begin{abstract}
We study the preparation and manipulation of states involving a
small number of interacting particles. By controlling the
splitting and fusing of potential wells, we show how to
interconvert Mott-insulator-like and trapped BEC-like states. We also
discuss the generation of ``Schr\"odinger cat" states by splitting
a microtrap and taking into practical consideration the asymmetry
between the resulting wells. These schemes can be used to perform
multiparticle interferometry with neutral atoms, where
interference effects can be observed only when all the
participating particles are measured.
\end{abstract}

\maketitle

Entanglement is at the root of Bell's theorem, which exposes the
differences between quantum theory and a local classical theory
based on elements of reality~\cite{bell64}. The predictions
of quantum mechanics have been experimentally
observed with entangled Einstein-Podolsky-Rosen (EPR)
pairs~\cite{epr_exp1,epr_exp2} as well as
Greenberger-Horne-Zeilinger (GHZ) triples~\cite{pan00}. A related
consequence of entanglement is the possibility of multiparticle
interferometry. Given a maximally entangled system of
$N$-particles (a ``Schr\"odinger cat" state) a measurement of
interference between different parts of the wave function 
corresponding to a single particle yields random results. It is
only when performing a coincidence measurement on all $N$
particles that an interference pattern is
revealed~\cite{greenberger93}. Experimental confirmation of this
result has been obtained using photonic EPR pairs~\cite{epr_exp2,
ghosh87} and internal states of four ions in the same
trap~\cite{sackett00} but no experiments have been performed using
a larger number of particles. The latest generation of experiments
with photons rely on parametric down-conversion, which has the
technical disadvantage of an exponentially decreasing number of
useful counts as $N$ increases.
Given that
entanglement is the key ingredient in all quantum computation and
quantum communication schemes, clean experimental studies of its
consequences have become an active topic of research in the last
decade.

In recent years several papers~\cite{jaksch99,andersson00}
have suggested the generation of entanglement between
neutral atoms confined in traps by using their interaction in
controlled atomic collisions. The atoms are guided in their motion
and their evolution yields the required entanglement of internal
states. Other schemes to achieve this sort of entanglement
starting from BECs have been suggested~\cite{bec_entanglement}. In
this letter we present two general $N$-atom nonlinear processes.
The first one is used to convert a Mott-insulator-like (MI)
state~\cite{Mott_insulator} into a state with all particles in the
(many-body) ground state of a single trap (BEC-like state); its
reverse process converts the BEC state into a MI state. The second
process is used to generate a Schr\"odinger cat state starting
from a BEC state by controlling the splitting of the well. As an
application of the processes we discuss a scheme for multiparticle
interferometry with spatially separated paths.

\begin{figure}[t]
\includegraphics[width=3in]{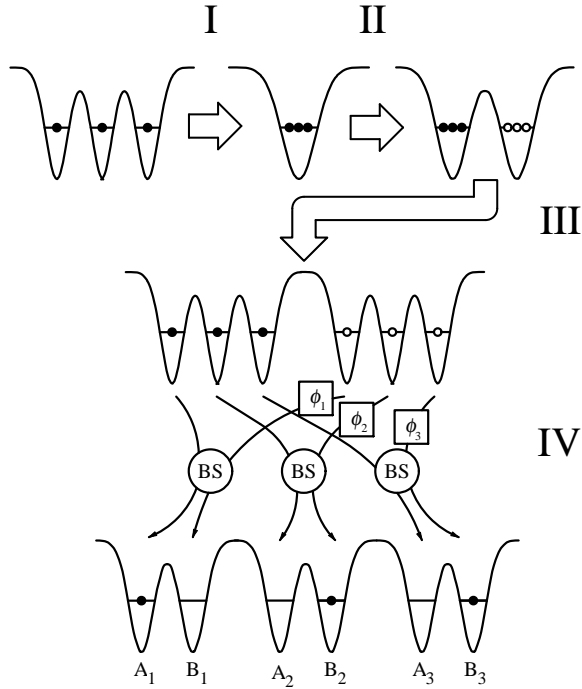}
\caption{\label{fig:pictorial} Schematics of the multiparticle
interferometry procedure. Stage I - creation of $N$ atoms in the
ground state of the trap starting with $N$ individual atoms in $N$
traps. Stage II - creation of ``Schr\"odinger cat'' state. Stage
III - spatial separation of the atoms. Stage IV - applying phases,
combining on the beamsplitters and measurement.}
\end{figure}

In the first process, which is also
stage I of the interferometry setup, we
start with a collection of $N$ atoms in the ground states of $N$
independent traps (MI state). These separate atoms can be extracted from a
reservoir using a quantum tweezer~\cite{opt_delivery} recently proposed
by our group. Alternatively, single atoms stored and detected in
micro-optical traps (which have been experimentally
reported~\cite{opt_detection} but are in excited states of the
trap) can be Raman cooled individually to the ground state.
%
The BEC state
(stage I, Fig.~\ref{fig:pictorial})
is achieved by bringing together the $N$ wells adiabatically
if the interaction between atoms is repulsive, as will be shown in
detail below. This is a consequence of the
quantum adiabatic theorem,
since the MI state is the ground state when the wells are far apart.
%
Our evolution is then represented by
\begin{equation}
|w_1\,w_2\,\ldots\,w_N\rangle \rightarrow |\Psi_I \rangle = |w\,w\,\ldots\,w\rangle,
\end{equation}
where the states are properly symmetrized bosonic states.

In the second process (stage II of multiparticle interferometry)
the interaction is switched to attractive. This can be done by
using a Feshbach resonance~\cite{feshbach_resonance}. Starting
from the BEC state, we slowly split the well into two
approximately equal mictrotraps, which we label as $L$ and $R$. The
lowest energy states are then the ones having all atoms in the
left or in the right well. Since initially the system is in the
ground state, by separating the traps at some slow rate $v$ when
the wells are far apart we get a linear combination of these two
nearly degenerate states, i.e. the system is in the Schr\"odinger
cat state
\begin{equation}
|\Psi_I\rangle \rightarrow |\Psi_{II}\rangle = \alpha |L L \ldots
L\rangle + \beta e^{i\theta} |R R \ldots R\rangle,
\end{equation}
with $\alpha$, $\beta$, and $\theta$ real. For perfectly symmetric
traps, $\alpha=\beta$ and $\theta=0$, but any asymmetry makes
these parameters rate dependent, as will be discussed in detail
below.

Additional processes are needed to realize multiparticle interferometry.
During stage III, the interaction is switched back to repulsive
and each of the two traps is separated to $N$. This stage can be
seen as the inverse of stage I applied to the wells $L$ and $R$.
Again, if the separation is done adiabatically the system remains
in the ground state which in this case corresponds to a single
atom in each one of the wells. The state is now
\begin{equation}
|\Psi_{II}\rangle \rightarrow \left| \Psi_{III} \right\rangle  =
\alpha \left| {L_1 L_2  \ldots L_N } \right\rangle  + \beta
e^{i\theta} \left| {R_1 R_2  \ldots R_N } \right\rangle.
\end{equation}
Subsequently, the atoms in wells derived from the original $R$ well
are subjected to additional phase shifts $\phi _1 ,\phi _2 \ldots
\phi _N$, which can be applied, for example, by adjusting the
depth of the wells adiabatically.

In the final, stage IV of the scheme, we combine states $L_i$ and
$R_i$ in a 50-50 beamsplitter~\cite{andersson99}. Notice that in
the experiment only one of these two is occupied so the
interatomic interaction plays no role in this stage. We denote the
outputs of each beamsplitter by $A_i$ and $B_i$ and assign a value
of $+1$ to the measurement of an atom in channel $A_i$, and $-1$
to the measurement of atom in channel $B_i$. The probability,
$P(+1)$, that the product of all measurements gives $+1$ (for
instance $A_1 B_2 B_3$ in the case of three atoms) is $\left( 1-
\alpha \beta \cos \left( {\Delta  + \theta } \right) \right) /2$,
where $\Delta  = \sum\limits_{i = 1}^N {\phi _i }$. The
probability for the product to be $-1$ is $P(-1) = 1 - P(+1)$,
hence the expectation value over a large number of measurements is
$- \alpha \beta \cos \left( {\Delta  + \theta  } \right)$. We
would like to stress that a correlated measurement of less that
$N$ atoms does not show any dependence on phase and appears
random.

\begin{figure}[t]
\includegraphics[width=3.25in]{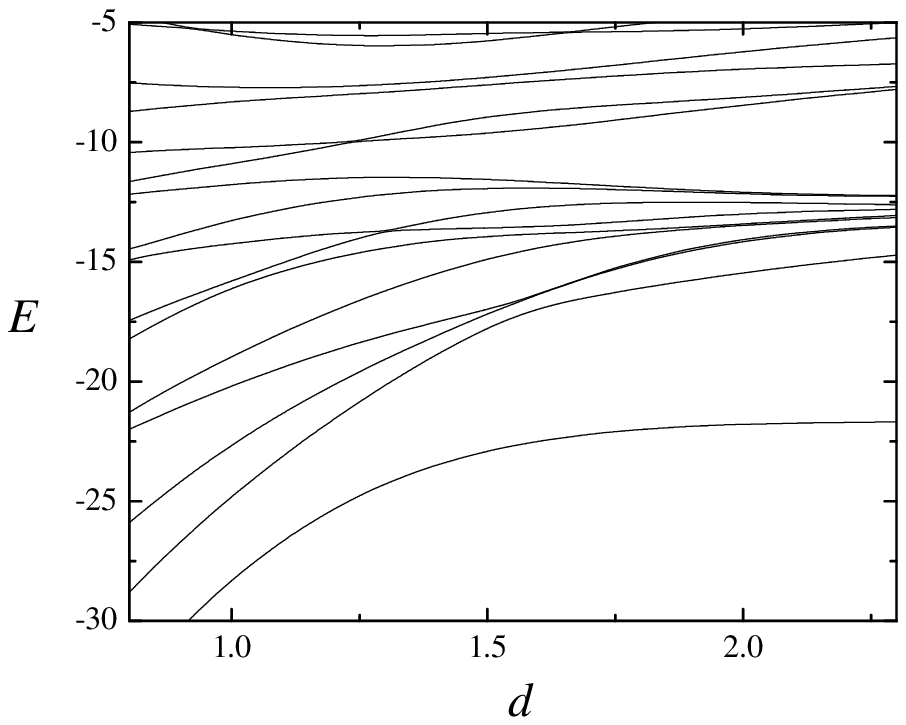}
\caption{\label{fig:levrep}  Stage I: adiabatic energy
levels for three atoms in three wells with repulsive interaction
as a function of $d$. The other parameters of the potential are
$V_0=10$, $\sigma=0.5$, $U_0=10$, $q_3 = -q_1 =10^{-4}$, $q_2 =
0$.}
\end{figure}



In order to obtain the relevant parameters for the operation of
our inteferometer, we study the evolution of an $N$ particle
system using optical microtraps. As an example, we numerically
solve the Schr\"odinger equation in the case of three atoms in a
quasi-1D configuration. This is achieved by strongly trapping the
atoms in the perpendicular dimensions, effectively freezing these
degrees of freedom. We scale the equations choosing units of
length $L_u  = {\rm 2~}\mu {\rm m}$, of energy $E_u = \hbar ^2
/\left( {2M_u L_u^2 } \right)$ and of time $t_u = \hbar /E_u$. The
particle interaction is represented by a delta-function potential
\begin{equation}
U(x_1 ,x_2 ) = U_0 \delta (x_1  - x_2 ).
\end{equation}
The atoms are also subject to external potentials due to the optical traps,
which in each stage are
\begin{equation}
\begin{array}{c}
V_{I, \,III}(x,d)= \sum\limits_{i = 1}^3 {(1 + q_i )} V(x,(i - 2)d),\\
V_{II} (x,d) =  (1+q_1) V  (x,-d/2) +(1+q_2) V  (x,+d/2),
\end{array}
\end{equation}
with
\begin{equation}
V(x,d) = - V_0 \exp \left( { - \frac{{\left( {x + d} \right)^2 }}
{{2\sigma ^2 }}} \right).
\end{equation}
The $q_i$ parametrize the asymmetry between the intensities of the
beams defining the different wells; we assume that these
are $10^{-4}$. 

\begin{figure}[t]
\includegraphics[width=3.25in]{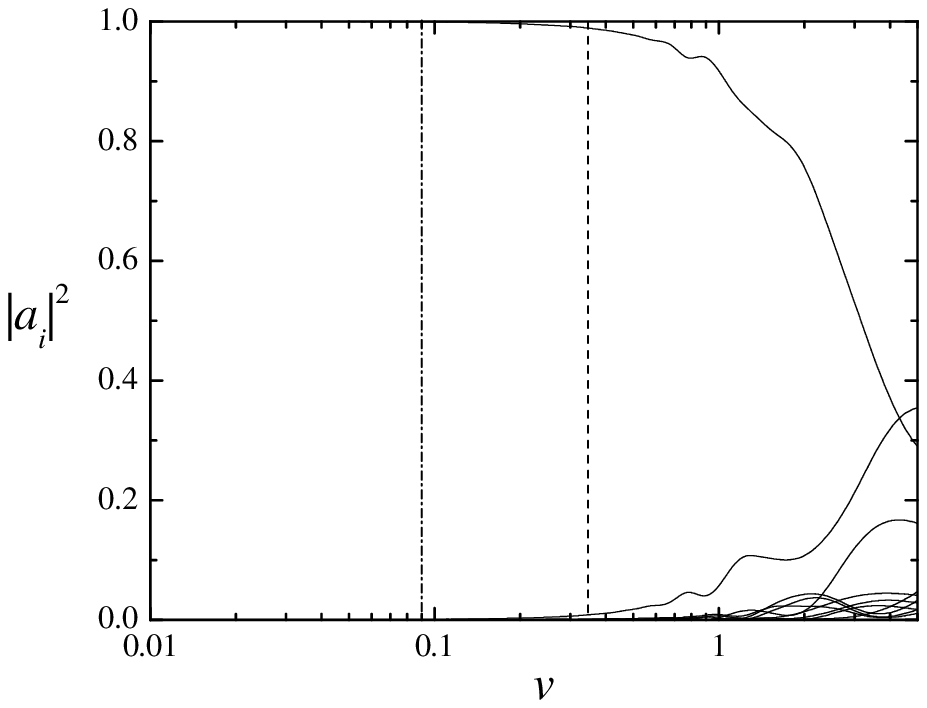}
\caption{\label{fig:dynrep} Stage III: probabilities to find the
system in the adiabatic states after a single well with three
atoms is split into three wells with an atom per well ($d_{\rm final}=3.0$) as a
function of the speed $v$. The energy levels are the ones shown in
Fig.~\ref{fig:levrep}. For velocities smaller than denoted with
dashed line probability to state in the ground state is larger
then 0.99, for velocities larger than denoted with dashed-dotted
line dephasing is less then 0.1. For stage I the dynamics are very
similar except there is no limit on how slow the process could be
done.}
\end{figure}

Let us consider first the evolution during the first and the third
stages of the operation. There are four different energy scales in
the problem. The first one is the energy difference between the
energy levels localized in different wells, which we can estimate
as $E_{\rm asym}\approx qV_0$. The second one is the energy
required to move one of the atoms to an already occupied well,
estimated to be $E_{\rm int} \approx U_0/\sigma_0$ where 
$\sigma_0 = (V_0/\sigma^2)^{1/4}$ is the width of the wave function 
in a well. The third scale
is the energy $E_{\rm exc} \approx \sigma_0^{-2}$ required to
put one of the atoms in an excited state of one of the traps. The
last energy scale ($E_D \approx (\pi/ND)^2$) is the energy
required to excite the atoms out of the ground state when the
distance between the wells is $D \approx 2\sigma$, at which time
the trap can be approximated by a square well of width $ND$. 
We operate in the regime in which
\begin{equation} \label{energy scales}
E_{\rm asym} \ll E_{\rm int}, \, E_{\rm exc}, \, E_D.
\end{equation}
Figure \ref{fig:levrep} shows the dependence of the adiabatic
levels on the separation $d$ during this stage. The presence of
the small asymmetry in this stage does not affect the nature of
the ground state, which is non-degenerate.
Joining or separating the wells at a slow speed keeps the system
in the ground state, i.e. the lowest curve in the figure. We can
estimate the rate at which the adiabaticity is lost by applying
the Landau-Zener formula~\cite{LZformula}, $ v_{\rm ad} \approx
(\Delta E_{\rm gap})^2 / (dE/dx)$. The slope can be estimated as
$\sqrt{NV_0}/\sigma^2$. The size of the gap depends on which of
the three large energy scales in (\ref{energy scales}) is the
smallest. In the example that we are presenting, all three are
roughly the same order of magnitude.  The probabilities $|a_i|^2 =
| \langle \psi_i|\psi \rangle |^2$ to find the system in the
states $|\psi_i\rangle$ at the end of the evolution is plotted in
Figure \ref{fig:dynrep} as a function of the speed $v$. In our
example, the critical rate is $v_{cI,2} = 0.35$; the probability
to find the system in other states is less than 0.01. For
multiparticle interferometry it is critical not to accumulate an
additional phase during the third process due to the asymmetry
between the right and left set of wells. This gives rise to a lower bound
for the allowed velocity, as explained below. For the parameters
chosen in the figure this is $v_{cI,1}=0.09$.

\begin{figure}[b]
\includegraphics[width=3.25in]{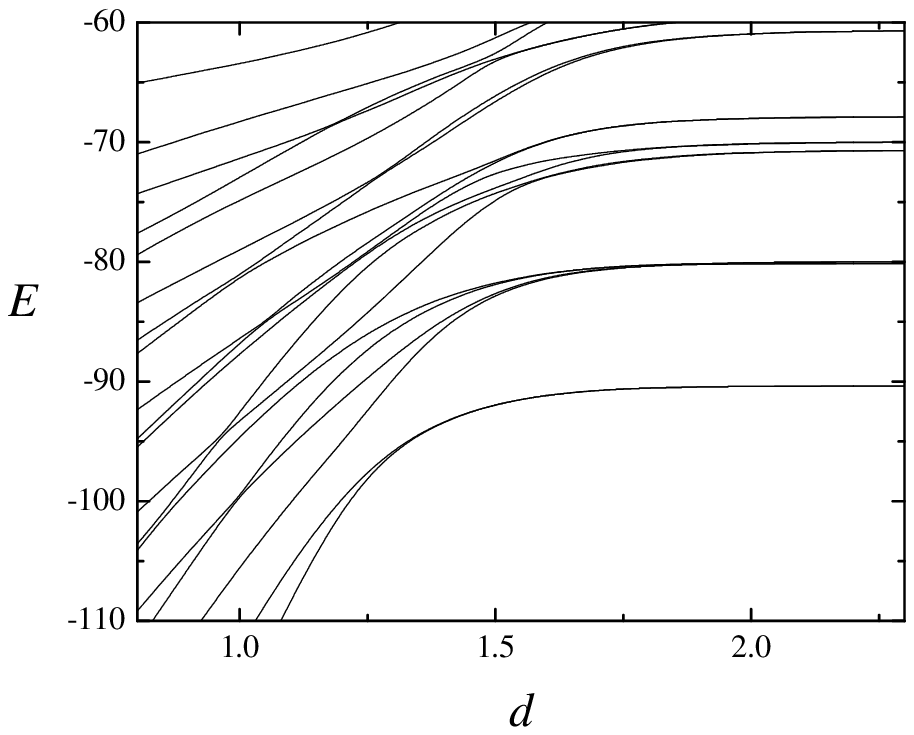}
\caption{\label{fig:levatt} Stage II: adiabatic levels of three
atoms in two wells in the case of attractive interaction for
different values of the separation $d$ . The other parameters of
the potential are $V_0=30$, $\sigma=0.5$, $U_0=-4$, $q_1=0$, $q_2
= 10^{-4}$.}
\end{figure}

Between these stages and stage I we need to change the sign of the
effective interaction between the particles. For the cases we are
considering, the particles remain the ground state with very high
probability (of the order of 99\%) even if this change is
performed suddenly.

During stage II the adiabatic energy levels as a function of $d$
are shown in Figure \ref{fig:levatt}. Once again we have four
energy scales, which can be approximated by $E_{\rm asym} \approx
NqV_0$, $E_{\rm int} \approx (N-1) |U_0|/\sigma_0$, 
$E_{\rm exc} \approx
\sigma_0^{-2}$,
 and $E_D \approx (\pi/2D)^2$.
Once again, we work in the regime in which (\ref{energy scales})
is valid. Separating the wells adiabatically maintains the system
in the ground state, which corresponds to all $N$ atoms being in
the lowest of the two wells, which is not the desired state. In
order to mix the lowest two energy states we need to evolve the
system non-adiabatically with respect to the lowest gap but at a
slow enough speed to remain adiabatic with respect to the larger
gap. Below $v_{cII,2}=0.27$ the probability
to tunnel to these excited states is less then 0.01 and
entanglement is obtained with $  \alpha \beta
= 0.99$ or larger. On the other hand, the asymmetry yields
a dephasing between the two parts of the wave function $\theta =
E_{\rm asym} t_{\rm sep}$, where the separation time is inversely
proportional to the velocity $v$. Allowing a maximum dephasing
$\phi_{\rm max}$, we must go faster than $v_{cI,1}\approx q
V_{0,III} N/ \phi_{\rm max} $. This calculation assumes, however, that the
asymmetry is constant. In a practical situation, $q$ is driven by
fluctuations in the laser power, and consequently the phase
$\theta$ grows diffusively, as the square root of $t_{\rm sep}$
instead of linearly, making the condition less restrictive.

\begin{figure}[t]
\includegraphics[width=3.25in]{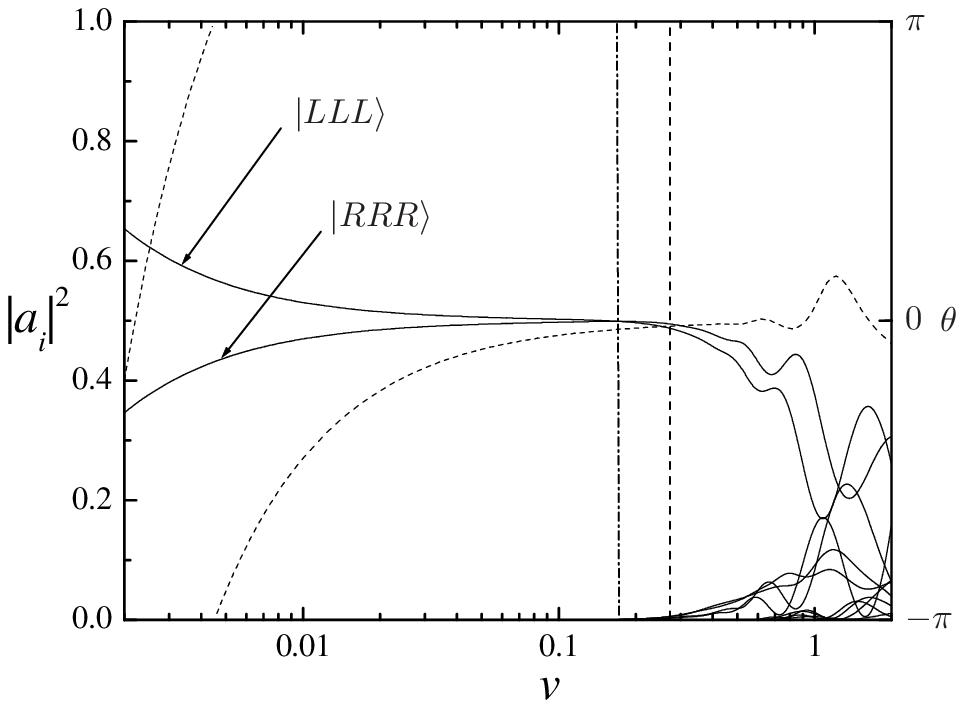}
\caption{\label{fig:dynatt} Stage II: full lines are probabilities
to find the system in the adiabatic states after the separation of
one well with three atoms to two ($d_{\rm final}=3.0$) as a function of the speed $v$.
The dashed line is $\theta$. The interaction is attractive and the
parameters are the ones used in Fig.~\ref{fig:levatt}. For
velocities in the interval between vertical lines the desired
state is prepared with probability of 0.99 and dephasing smaller
than 0.1.}
\end{figure}

The only two conditions for the applicability of the method are
related to the asymmetry of the potential. As long as condition
(\ref{energy scales}) is met and as long as $v_{c,2}$ is larger
than $v_{c,1}$, there is a range of velocities for which the
operation is successful. The critical velocities have different
dependence on $N$, so for fixed values of the parameters defining
the potential and the interaction, there is a largest number of
atoms for which this happens. However, by choosing a different set
of parameters this condition can be relaxed. In particular, the
strong $N^{-2}$ dependence of the preparation of the MI state can
be overcome by separating the atoms in series instead of doing it
parallel (for $N=2^n$, we can think of $n$ steps in which each
well is split into two).

Finally, numerical values for realistic experimental parameters
are given. In the model described above the effective interaction
between atoms is determined by the scattering length $a$ and the
strength of the confinement in the transverse direction. The
frequency $\omega _ \bot$ of the confinement in the case in which
the system stays in the ground state of transverse motion may be
expressed in terms of the dimensionless interaction parameter
$U_0$ used above~\cite{jaksch99} as
\begin{equation}
\omega _ \bot   = \frac{{U_0 \hbar }}{{4\left|a\right|M_u L_u }}.
\end{equation}
Hence it is desirable to use atoms with the largest product of
mass and scattering length possible. In Table~\ref{tab:par} we
present the rescaled values used in the calculation for two
workhorses of cold atom experiments, sodium and rubidium. The
magnetic fields needed to observe Feshbach resonances in alkali
atoms are typically hundreds of gauss \cite{feshbach_resonance}.
In the proposed scheme for multiparticle interferometry one should
work on the side of the resonance where the scattering length
changes sign to avoid the losses associated with crossing the
resonance.


We acknowledge support from the NSF, the R.A. Welch Foundation,
MGR also acknowledges support from the Sid W. Richardson Foundation.


\begin{table}[h]
\caption{\label{tab:par}Parameters of the numerical estimates in
dimensional units. For the estimates we take scattering lengths of
${}^{23}{\rm Na}$ $a_t = 65 a_0$ and ${}^{87}{\rm Rb}$ $a_t = 106
a_0$ in triplet states with no magnetic field with $a_0$ being the
Bohr radius~\cite{abeelen99}; we assume that near a Feshbach
resonance the values will be of the same order of magnitude.
}
\begin{ruledtabular}
\begin{tabular}{cccc}
Parameter&${\rm Na}$&${\rm Rb}$& Units\\
\hline
$\omega _ \bot$ & 79.9 & 13.0 & 2$\pi$ kHz\\
$v _ {cI,1}$ & 62.2 & 16.5 & $\mu$m/s\\
$v _ {cI,2}$  & 242 & 64.0 & $\mu$m/s\\
$v _ {cII,1}$  & 117 & 31.1 & $\mu$m/s\\
$v _ {cII,2}$  & 186 & 49.4 & $\mu$m/s\\
$V_{0,II}$ & 2.47 & 0.665 & $ h \times$kHz\\
\end{tabular}
\end{ruledtabular}
\end{table}

\end{document}